# Magnetic excitations in the spin-trimer compounds Ca$_3$Cu$_{3-x}$Ni$_x$(PO$_4$)$_4$ (x=0,1,2)

A. Podlesnyak[1,2], V. Pomjakushin[1], E. Pomjakushina[1,3], K. Conder[3], and A. Furrer[1]

[1] Laboratory for Neutron Scattering, ETH Zurich & PSI Villigen, CH-5232 Villigen PSI, Switzerland

[2] Hahn-Meitner-Institut, SF-2, Glienicker Str. 100, D-14109 Berlin, Germany

[3] Laboratory for Developments and Methods, PSI Villigen, CH-5232 Villigen PSI, Switzerland

**Abstract:**

Inelastic neutron scattering experiments were performed for the spin-trimer compounds Ca$_3$Cu$_{3-x}$Ni$_x$(PO$_4$)$_4$ (x=0,1,2) in order to study the dynamic magnetic properties. The observed excitations can be associated with transitions between the low-lying electronic states of linear Cu-Cu-Cu, Cu-Cu-Ni, and Ni-Cu-Ni trimers which are the basic constituents of the title compounds. The exchange interactions within the trimers are well described by the Heisenberg model with dominant antiferromagnetic nearest-neighbor interactions J. For x=0 we find J$_{Cu-Cu}$=(-4.74±0.02) meV which is enhanced for x=1 to J$_{Cu-Cu}$=(-4.92±0.06) meV. For x=1 and x=2 we find J$_{Cu-Ni}$=(-0.85±0.10) meV and an axial single-ion anisotropy parameter D$_{Ni}$=(-0.7±0.1) meV. While the x=0 and x=1 compounds do not exhibit long-range magnetic ordering down to 1 K, the x=2 compound shows antiferromagnetic ordering below T$_N$=20 K, which is compatible with the molecular-field parameter λ=(0.63±0.12) meV derived by neutron spectroscopy.

# 1. Introduction

Gapped quantum magnets containing $S=\frac{1}{2}$ $Cu^{2+}$ ions are of particular interest due to the competition between the classical crystalline and the quantum Bose-Einstein condensed ground states of bosons. Gapped ground states naturally arise from dimer systems in which the two $Cu^{2+}$ ions are antiferromagnetically coupled to form a spin singlet, separated by an energy gap from the excited spin triplet. $Cu^{2+}$ dimer systems are realized e.g. for $SrCu_2O_3$ [1], $SrCu_2(BO_3)_2$ [2], $BaCuSi_2O_6$ [3,4], $TlCuCl_3$ [5], and $Cs_2CuCl_4$ [6]; $TlCuCl_3$ is a particularly nice example in which the field- and pressure-induced Bose-Einstein condensation of bosonic triplet states was directly verified by inelastic neutron scattering experiments [7,8]. On the other hand, examples of spin trimer systems are rare in nature. The few compounds known to date include $La_4Cu_3Mo_3O_{12}$ in which the three $Cu^{2+}$ spins form a triangle, so that the antiferromagnetic intra-trimer interactions are frustrated [9], and $Sr_2Cu_3O_5$ containing three-leg ladders whose suceptibility data, however, reflect a gapless spin excitation spectrum [1]. $A_3Cu_3(PO_4)_4$ (A=Ca,Sr,Pb) is a novel spin trimer system in which the intertrimer interaction is weak, so that the antiferromagnetically coupled $Cu^{2+}$ spins give rise to a doublet ground state [10]. The doublet ground-state can eventually be changed to a singlet ground-state by substituting a $Cu^{2+}$ spin in the trimer by $Ni^{2+}$ (S=1), which would offer for the first time the realization of a gapped quantum spin trimer system. The present work was motivated by this hypothesis which, however, was not met due to the particular features of the mixed Cu-Ni trimers. Nevertheless, it is expected that the present work encourages the further search for gapped spin trimer systems.

The magnetic properties of the compounds $A_3Cu_3(PO_4)_4$ (A=Ca,Sr,Pb) were studied by magnetic susceptibility, heat capacity, and magnetization experiments. Long-range magnetic ordering was observed below $T_N=0.8$ K, 0.9 K, and 1.3 K for A=Ca, Sr, and Pb, respectively [11,12]. From inelastic neutron scattering experiments the intra-trimer interactions were determined and found to be antiferromagnetic [13]. The structural and static magnetic properties of the mixed compounds $Ca_3Cu_{3-x}Ni_x(PO_4)_4$ (x=0,1,2) were investigated by neutron diffraction and magnetic susceptibility experiments [14]. The x=0 compound crystallizes in a monoclinic structure (space group $P2_1/a$, N14) with the cell parameters a=17.621 Å, b=4.902 Å, c=8.922 Å, and β=124.07° at room temperature. The $Cu^{2+}$ ions occupy the two crystallographic positions (2a) in the middle and (4e) at the ends of the linear Cu-Cu-Cu trimer. The x=1 compound crystallizes in the same space group. The occupation of the trimer is such that the middle position (2a) is exclusively occupied by $Cu^{2+}$, whereas the end positions (4e) are statistically populated with $Cu^{2+}$ and $Ni^{2+}$, thus there are three different types of trimers: Cu-Cu-Cu, Cu-Cu-Ni, and Ni-Cu-Ni. Magnetic susceptibility measurements indicate the absence of long-range magnetic ordering for T>2 K. The x=2 compound crystallizes in the space group C2/c (N15) with doubled unit cell along the c-axis, and only trimers of the type Ni-Cu-Ni are present. Below $T_N$=20 K antiferromagnetic ordering with two propagation vectors $\mathbf{k_1}=[\frac{1}{2},\frac{1}{2},0]$ and $\mathbf{k_2}=[-\frac{1}{2},\frac{1}{2},0]$ was found, i.e., the Ni-Cu-Ni trimers form two different sublattices a and b in which the $Cu^{2+}$ moments are oriented parallel and antiparallel to the $Ni^{2+}$ moments, respectively. The individual magnetic moments amount to $\mu_{Ni}$=1.89(1) $\mu_B$ and $\mu_{Cu}$=0.62(2) $\mu_B$ at low temperatures, i.e., they are not fully saturated.

The information collected in Ref. [14] was an important prerequisite for the inelastic neutron scattering study of the dynamic magnetic properties of the compounds $Ca_3Cu_{3-x}Ni_x(PO_4)_4$ (x=0,1,2) described in the present work which is organized as follows. Section 2 provides a short summary of the experimental details. The theoretical background is described in Section 3 in which the energy states of different trimer types as well as the neutron cross-section for transitions between these states are derived. The main part of the paper is Section 4 which presents the experimental data as well as their analyses in terms of the quantum spin models outlined in the preceding Section. Finally the results are discussed and some conclusions are given in Section 5.

## 2. Experimental

Polycrystalline samples of $Ca_3Cu_{3-x}Ni_x(PO_4)_4$ (x=0,1,2) were synthesized by a solid state reaction as proposed in Ref. [10] using CuO, NiO, $CaCO_3$ and $NH_4H_2PO_4$ of a minimum purity of 99.99%. The respective amounts of the starting reagents were mixed and heated very slowly up to 600ºC amd then annealed at 900ºC during at least 100 h, with several intermediate grindings in alumina crucibles. All the samples were characterized prior to the present experiments by susceptibility and neutron powder diffraction experiments as described in detail in Ref. [14].

The inelastic neutron scattering experiments were carried out for polycrystalline samples of $Ca_3Cu_{3-x}Ni_x(PO_4)_4$ (x=0,1,2) with use of the high-resolution time-of-flight spectrometer FOCUS at the spallation neutron source SINQ at PSI Villigen. The measurements were performed with incoming neutron energies $E_i$ = 20, 15, 4.4, and 2.7 meV, giving energy resolutions at the elastic position of 1.68, 1.08, 0.18, and 0.06 meV, respectively. The energy resolutions are gradually improving with increasing energy transfer on the energy-loss side of the spectrum, typically by 10% at energy transfers corresponding to half the incoming neutron energy. The scattered neutrons were detected by an array of $^3$He counters covering a large range of scattering angles 10º≤Φ≤130º. The data from different detectors were combined to form groups with constant modulus of the scattering vector **Q**. For the data presentation we use the notation <Q> to indicate that the Q-width was enlarged by ±ΔQ (of the order of a few tenths of Å$^{-1}$) to achieve a sufficient counting statistics. The samples were enclosed in Al cylinders (diameter 12 mm, height 45 mm) and placed into a He cryostat to reach temperatures T≥1.5 K. Additional experiments were performed for the empty container as well as for vanadium to allow the correction of the raw data with respect to background, detector efficiency, absorption, and detailed balance effects according to standard procedures.

## 3. Theoretical background

We consider homogeneous spin trimers with identical ions as well as mixed spin trimers containing two different ions A and B. This gives rise to three different trimer types I, II, and III as visualized in Figure 1. In the classification of the trimer types we add a further index A or B in order to identify the majority ion. In the following Sections we derive the energy states of the different trimer types as well as the neutron cross-section for transitions between these states.

### 3.1. Homogeneous spin trimers of type I

The Heisenberg exchange Hamiltonian of a trimer of type I.A and I.B is described by

$$H_{ex} = -2[J(\mathbf{S_1}\cdot\mathbf{S_2}+\mathbf{S_2}\cdot\mathbf{S_3}) + J'\mathbf{S_1}\cdot\mathbf{S_3}] \;, \tag{1}$$

where J and J' denote the nearest-neighbor and next-nearest-neighbor exchange parameters, respectively, and the $\mathbf{S_i}$ stand for the spin operators. The spin quantum number of an individual ion is $S_i$. For a complete characterization of the trimer states we need additional quantum numbers $S_{13}$ and S resulting from the vector sums $\mathbf{S_{13}}=\mathbf{S_1}+\mathbf{S_3}$ and $\mathbf{S}=\mathbf{S_1}+\mathbf{S_2}+\mathbf{S_3}$ with $0 \leq S_{13} \leq 2S_i$ and $|S_{13}-S_i| \leq S \leq |S_{13}+S_i|$, respectively. The trimer states are therefore defined by the wave functions $|S_{13},S>$, and their degeneracy is (2S+1). With this choice of spin quantum numbers the Hamiltonian (1) is diagonal, thus the energy eigenvalues can readily be derived:

$$E(S_{13},S) = -J[S(S+1)-S_{13}(S_{13}+1)-S_i(S_i+1)] - J'[S_{13}(S_{13}+1)-2S_i(S_i+1)] \ . \qquad (2)$$

Application of Eq. (2) to the compound $Ca_3Cu_3(PO_4)_4$, i.e., $A=Cu^{2+}$ and $S_i=\frac{1}{2}$, yields the following eigenstates:

$$\begin{aligned}
|0,\tfrac{1}{2}> \ &: \ E(0,\tfrac{1}{2}) \ = \ \phantom{2J -\ } \tfrac{3}{2}J' \\
|1,\tfrac{1}{2}> \ &: \ E(1,\tfrac{1}{2}) \ = \ 2J - \tfrac{1}{2}J' \\
|1,\tfrac{3}{2}> \ &: \ E(1,\tfrac{3}{2}) \ = \ -J - \tfrac{1}{2}J'
\end{aligned} \qquad (3)$$

The ground state is the quartet $|1,\tfrac{3}{2}>$ for ferromagnetic coupling J>0 and the doublet $|1,\tfrac{1}{2}>$ for antiferromagnetic coupling J<0, and the overall trimer splitting is exactly 3J for |J|>|J'|.

### 3.2. Mixed spin trimers of type II

The exchange Hamiltonian of a mixed spin trimer of type II.A is identical to Eq. (1). However, the resulting eigenvalues depend now on the detailed spin quantum numbers $S_i^A$ and $S_i^B$ of the ions A and B, respectively:

$$E(S_{13},S) = -J[S(S+1)-S_{13}(S_{13}+1)-S_i^B(S_i^B+1)]-J'[S_{13}(S_{13}+1)-2S_i^A(S_i^A+1)] \quad . \quad (4)$$

The spin quantum numbers follow the rules $0 \leq S_{13} \leq 2S_i^A$ and $|S_{13}-S_i^B| \leq S \leq |S_{13}+S_i^B|$.

Application to A=$Cu^{2+}$ with $S_i^A=\frac{1}{2}$ and B=$Ni^{2+}$ with $S_i^B=1$ yields the following eigenstates:

$$\begin{aligned}
|0,1\rangle &: E(0,1) = \tfrac{3}{2}J' \\
|1,0\rangle &: E(1,0) = 4J - \tfrac{1}{2}J' \\
|1,1\rangle &: E(1,1) = 2J - \tfrac{1}{2}J' \\
|1,2\rangle &: E(1,2) = -2J - \tfrac{1}{2}J'
\end{aligned} \quad (5)$$

For antiferromagnetic coupling J<0 the singlet |1,0> is the ground state. The realization of a singlet ground state offers the observation of a field-induced quantum phase transition and eventually the occurrence of a Bose-Einstein condensation, which so far has not yet been achieved for a spin trimer system. As mentioned in the Introduction, this particular property was actually the motivation for the present work, however, spin trimers of type II.A are unfortunately not realized in the compound $Ca_3Cu_2Ni(PO_4)_4$ [14].

The reverse application, i.e., a mixed spin trimer of type II.B by choosing A=$Ni^{2+}$ and B=$Cu^{2+}$ yields the following eigenstates:

$$\begin{aligned}
|0,\tfrac{1}{2}\rangle &: E(0,\tfrac{1}{2}) = 4J' \\
|1,\tfrac{1}{2}\rangle &: E(1,\tfrac{1}{2}) = 2J + 2J' \\
|1,\tfrac{3}{2}\rangle &: E(1,\tfrac{3}{2}) = -J + 2J' \\
|2,\tfrac{3}{2}\rangle &: E(2,\tfrac{3}{2}) = 3J - 2J' \\
|2,\tfrac{5}{2}\rangle &: E(2,\tfrac{5}{2}) = -2J - 2J'
\end{aligned} \quad (6)$$

For antiferromagnetic coupling J<0 the quartet $|2,\frac{3}{2}\rangle$ is the ground state, and the overall trimer splitting amounts to exactly 5J for |J|>|J'|.

### 3.3. Mixed spin trimers of type III

The exchange Hamiltonian of mixed spin trimers of type III.A and III.B is given by

$$H_{ex} = -2\,[\,J_{12}\,\mathbf{S_1}\cdot\mathbf{S_2} + J_{23}\,\mathbf{S_2}\cdot\mathbf{S_3} + J'\,\mathbf{S_1}\cdot\mathbf{S_3}\,]\,, \qquad (7)$$

which is no longer diagonal as for the trimers of type I and II. The eigenvalue problem can most conveniently be solved by using irreducible tensor techniques as applied for spin trimers e.g. by Griffith [15]. We apply this technique to trimers of type III.A with A=$Cu^{2+}$ and B=$Ni^{2+}$ which is relevant for the compound $Ca_3Cu_2Ni(PO_4)_4$. The basis spin states are those of Eq. (5), however, $|S_{13},S\rangle$ should be replaced by $|S_{12},S\rangle$, since it is convenient to couple $\mathbf{S_1}+\mathbf{S_2}=\mathbf{S_{12}}$ rather than $\mathbf{S_1}+\mathbf{S_3}=\mathbf{S_{13}}$. The off-diagonal terms of the Hamiltonian (7) result in a mixing of the states with equal spin quantum numbers S and M, i.e., for the triplet states $|0,1\rangle$ and $|1,1\rangle$. The results are displayed in Fig. 2 for different values of the dominating nearest-neighbor exchange parameters $J_{12}$ and $J_{23}$, whereas the next-nearest neighbor exchange is neglected, i.e., J'=0. For antiferromagnetic coupling $J_{12}<0$ the triplet $a_2|0,1\rangle - a_1|1,1\rangle$ is the ground state.

### 3.4. Anisotropy effects

For non S-state ions the isotropic exchange Hamiltonian discussed so far has to be extended to include an anisotropy term which we express by the single-ion Hamiltonian

$$H_{an} = D \sum_i \left(S_i^z\right)^2 . \tag{8}$$

The sum runs over all the non S-state ions of the trimer, i.e., for $Ni^{2+}$ in the compounds $Ca_3Cu_{3-x}Ni_x(PO_4)_4$. The total spin Hamiltonian is therefore $H_{ex}+H_{an}$. The anisotropy term has the effect to split the spin states $|S_{13},S\rangle$ into the states $|S_{13},S,\pm M\rangle$ where the additional spin quantum number M is defined by $-S \leq M \leq S$. Moreover, the Hamiltonian (8) gives rise to off-diagonal terms which produce a mixing of the states with equal spin quantum number M, e.g., for the triplet states $|0,1,\pm 1\rangle$, $|1,1,\pm 1\rangle$, and $|1,2,\pm 1\rangle$ in spin trimers of type III.A.

### 3.5. Magnetic field effects

The Hamiltonian

$$H_H = -g\mu_B H \sum_i S_i^z \tag{9}$$

describes the effect of a magnetic field H (along the z-axis) on the spin states $|S_{13},S,\pm M\rangle$ whose two-fold degeneracy is completely lifted to produce two singlets $|S_{13},S,M\rangle$ and $|S_{13},S,-M\rangle$ separated by the Zeeman energy $2Mg\mu_B H$, where g is the g-factor and $\mu_B$ the Bohr magneton. The magnetic field can be either an external field or the molecular field in the magnetically ordered state or both. The molecular field $H_{mf}$ is defined by

$$g\mu_B H_{mf} = \lambda <S> ,  \qquad (10)$$

where $\lambda$ is the molecular-field parameter and $<S>$ the expectation value of the total spin. The spin Hamiltonian in the magnetically ordered state is therefore given by $H_{ex}+H_{an}+H_H$. The field-induced level splitting is sketched in Fig. 3. In the presence of a molecular field $H_{mf}$ the energy $\Delta$ of the trimer transition $|S,M> \rightarrow |S',M'>$ changes to

$$\Delta_{mf} = \Delta + (M-M')g\mu_B H_{mf} \quad . \qquad (11)$$

### 3.6. Neutron cross-section

Basic formulae for the neutron cross-section of polynuclear spin clusters were given in Ref. [16]. We apply these formulae to spin trimers of type I and II. For polycrystalline material we find the following neutron cross-section for a transition from the initial state $|S_{13},S,M>$ to the final state $|S'_{13},S',M'>$:

$$\frac{d^2\sigma}{d\Omega d\omega} = C \left\{ \left[ 1+(-1)^{S_{13}-S'_{13}} \frac{\sin(2QR)}{2QR} \right] T_1^2 + \delta_{S_{13},S'_{13}} T_3^2 - 4\delta_{S_{13},S'_{13}} \frac{\sin(QR)}{QR} |T_1||T_3| \right\}$$

with  (12)

$$C = N \left( \frac{\gamma e^2}{m_e c^2} \right)^2 p \frac{k'}{k} F^2(Q) \exp\{-2W(Q)\} \delta\{\hbar\omega + E(S_{13},S,M) - E(S'_{13},S',M')\} .$$

N is the total number of spin trimers in the sample, p the population of the initial state $|S_{13},S,M\rangle$ governed by Boltzmann statistics, k and k' the wave numbers of the incoming and scattered neutrons, respectively, $\mathbf{Q}=\mathbf{k}-\mathbf{k'}$ the scattering vector, F(Q) the magnetic form factor, $\exp\{-2W(Q)\}$ the Debye-Waller factor, $\hbar\omega$ the energy transfer, $E(S_{13},S,M)$ and $E(S'_{13},S',M')$ the energy of the initial and final state, respectively, R the distance between the center spin (at site 2) and the end-standing spins (at sites 1 and 3), and $T_1$ and $T_3$ transition matrix elements defined in Ref. [16]. The remaining symbols have their usual meaning.

The transition matrix elements $T_1$ and $T_3$ carry essential information to derive the selection rules for spin trimers:

$$\Delta S_{13} = S_{13} - S'_{13} = 0, \pm 1 \,, \quad \Delta S = S - S' = 0, \pm 1 \,, \quad \Delta M = M - M' = 0, \pm 1 \,. \tag{13}$$

For the proper identification of particular trimer transitions the observation of the Q-dependence of the intensities is most helpful, since the cross-section formula (12) depends markedly on $\Delta S_{13}$ as visualized in Fig. 4 for spin trimers of type I.A.

For spin trimers of type III the cross-section formula (12) can be applied by changing the spin coupling scheme from $|S_{13},S\rangle$ to $|S_{12},S\rangle$ as described in Section 3.3.

## 4. Results and data analysis

### 4.1. A qualitative overview for $Ca_3Cu_{3-x}Ni_x(PO_4)_4$ (x=0,1,2)

Energy spectra of neutrons scattered from $Ca_3Cu_{3-x}Ni_x(PO_4)_4$ (x=0,1,2) at T=1.5 K and $\langle Q \rangle$=3 Å$^{-1}$ are shown in Fig. 5. By going from x=0 to x=2, i.e. the observed energy spectra exhibit two major differences. Firstly, the increase of the Ni content produces a considerable enhancement of the phonon scattering due to the large incoherent cross section of nickel, which is very pronounced e.g. at the high-energy tail of the x=2 spectrum. Secondly, the overall energy splitting diminishes from about 14 meV for x=0 to about 6 meV for x=2. The x=0 and x=2 spectra result from excitations associated with the Cu-Cu-Cu and Ni-Cu-Ni trimers, respectively. According to Eqs. (3) and (6) the overall energy splittings for Cu-Cu-Cu and Ni-Cu-Ni trimers are $3|J_{Cu-Cu}| \approx 14$ meV and $5|J_{Cu-Ni}| \approx 6$ meV, respectively, thus we can immediately conclude that $|J_{Cu-Cu}| \approx 4.7$ meV and $|J_{Cu-Ni}| \approx 1.2$ meV, i.e., $|J_{Cu-Cu}| \gg |J_{Cu-Ni}|$.

## 4.2. $Ca_3Cu_3(PO_4)_4$

Energy spectra of neutrons scattered from $Ca_3Cu_3(PO_4)_4$ at T=1.5 K are shown in Fig. 6. For small moduli of the scattering vector **Q** there are two well defined peaks superimposed on a low and constant background. With increasing Q the intensity of the two peaks is decreasing, whereas the background markedly raises over the whole energy spectrum due to phonon scattering. A Gaussian least-squares fit was applied to the observed energy spectra to yield the peak positions at energies $\varepsilon_1$=(9.44±0.03) meV and $\varepsilon_2$=(14.22±0.02) meV. The linewidth of both peaks corresponds to the instrumental energy resolution. The intensity behavior of the two inelastic lines is characteristic of magnetic excitations, thus we identify them as transitions out of the trimer ground-state to excited trimer states at $\varepsilon_1$ and $\varepsilon_2$. According to Eq. (3) the ratio $\varepsilon_1/\varepsilon_2 \approx 2/3$ can only be achieved for antiferromagnetic nearest-neighbor coupling J<0, thus the state $|1,\frac{1}{2}>$ is the ground state, and the peaks at $\varepsilon_1$ and $\varepsilon_2$ correspond to the excited trimer states $|0,\frac{1}{2}>$ and $|1,\frac{3}{2}>$, respectively. This identification is nicely confirmed by the Q-dependence of the intensities as shown in Fig. 4. Analyzing the observed trimer splittings according to Eq. (3) yields the coupling parameters

$J_{Cu-Cu} = (-4.74 \pm 0.02)$ meV , $J'_{Cu-Cu} = (-0.02 \pm 0.03)$ meV ,

which are in good agreement with the results published by Matsuda et al. [13]. Next-nearest-neighbor exchange coupling is clearly negligible.

### 4.3. $Ca_3Cu_2Ni(PO_4)_4$

As mentioned in the Introduction, three different trimer types are present for the compound $Ca_3Cu_2Ni(PO_4)_4$, thus the neutron spectroscopy data correspond to a superposition of the energy spectra of the trimers Cu-Cu-Cu, Cu-Cu-Ni, and Ni-Cu-Ni. An example of the observed energy spectra is shown in Fig. 5 which is characterized by two broad peaks centered around 5 meV and 9 meV, followed by a peak at $\varepsilon=(14.76\pm0.14)$ meV as determined by a Gaussian least-squares fit. The latter peak has to be attributed to the $|1,\frac{1}{2}>\rightarrow|1,\frac{3}{2}>$ transition of the Cu-Cu-Cu trimer, whose overall energy splitting corresponding to $\varepsilon=3|J_{Cu-Cu}|$ exceeds that of the Cu-Cu-Ni and Ni-Cu-Ni trimers as mentioned in Section 4.1. $|J_{Cu-Cu}|=\varepsilon/3$ is enhanced in the Ni-doped compound $Ca_3Cu_2Ni(PO_4)_4$ to $J_{Cu-Cu}=(-4.92\pm0.05)$ meV as compared to $J_{Cu-Cu}=(-4.74\pm0.02)$ meV derived for the Ni-free compound $Ca_3Cu_3(PO_4)_4$, see Section 4.2.

Fig. 7 displays data taken at low energies with increased energy resolution which exhibits three inelastic lines at energies $\varepsilon_1=(0.52\pm0.06)$ meV, $\varepsilon_2=(0.80\pm0.04)$ meV, and $\varepsilon_3=(1.05\pm0.06)$ meV as refined by a Gaussian least-squares fit. These lines have to be associated with excitations of the Cu-Cu-Ni and Ni-Cu-Ni trimers, since there are no low-lying excitations for the Cu-Cu-Cu trimer. A proper identification of these lines can be achieved by detailed considerations of their intensities $I_i$ which turn out to be $(I_1/I_2)_{obs}=(0.35\pm0.04)$, $(I_1/I_3)_{obs}=(0.47\pm0.06)$, and $(I_2/I_3)_{obs}=(1.35\pm0.11)$. The peak identification should then be confirmed by calculations based on the three model parameters $J_{Cu-Cu}$, $J_{Cu-Ni}$ and $D_{Ni}$ which have to be identical for both trimers. $J_{Cu-Cu}=-4.92$ meV is fixed (see preceding paragraph), thus we are left with two disposable parameters $J_{Cu-Ni}$ and $D_{Ni}$.

First we consider the low-lying excitations of the trimer Cu-Cu-Ni. As discussed in Section 4.1 we expect $|J_{23}|=|J_{Cu-Ni}| \ll |J_{12}|=|J_{Cu-Cu}|$. The spin state correlation diagram of Fig. 2 then suggests that the excited states are separated from the ground state approximately by $2|J_{12}|\approx 10$ meV, thus low-lying excitations can only result from the splitting of the ground-state triplet $a_2|0,1\rangle - a_1|1,1\rangle$ due to the single-ion anisotropy defined by Eq. (8). For D<0 or D>0 we have the doublet $a_2|0,1,\pm 1\rangle - a_1|1,1,\pm 1\rangle$ below or above the singlet $a_2|0,1,0\rangle - a_1|1,1,0\rangle$, respectively. Therefore one of the three inelastic lines in Fig. 7 has to be identified by the singlet-doublet ground-state splitting of the Cu-Cu-Ni trimer.

Similar considerations can be made for the trimer Ni-Cu-Ni. There are three low-lying excitations for $J_{Cu-Ni}>0$, whereas for $J_{Cu-Ni}<0$ only two low-lying excitations are expected, thus we can discard the possibility $J_{Cu-Ni}>0$. According to Eq. (6) the quartet $|2,\frac{3}{2}\rangle$ is the ground state for antiferromagnetic coupling $J_{Cu-Ni}$. The single-ion anisotropy raises the four-fold degeneracy of the ground state $|2,\frac{3}{2}\rangle$ to produce the doublet $|2,\frac{3}{2},\pm\frac{3}{2}\rangle$ below or above the doublet $|2,\frac{3}{2},\pm\frac{1}{2}\rangle$ for D<0 or D>0, respectively. In addition, with $|J_{Cu-Ni}|\approx 1.2$ meV the ground-state transition to the excited state $|1,\frac{1}{2},\pm\frac{1}{2}\rangle$ (which cannot be split by the single-ion anisotropy) is expected to fall into the low-energy window of the energy spectrum as well. With use of Eq. (12) we calculate the intensity ratio of these two excitations for both D>0 and D<0:

D>0: $I(|2,\frac{3}{2},\pm\frac{1}{2}> \rightarrow |2,\frac{3}{2},\pm\frac{3}{2}>) / I(|2,\frac{3}{2},\pm\frac{1}{2}> \rightarrow |1,\frac{1}{2},\pm\frac{1}{2}>) = 0.66$ ,

D<0: $I(|2,\frac{3}{2},\pm\frac{3}{2}> \rightarrow |2,\frac{3}{2},\pm\frac{1}{2}>) / I(|2,\frac{3}{2},\pm\frac{3}{2}> \rightarrow |1,\frac{1}{2},\pm\frac{1}{2}>) = 0.44$ .

The intensity ratio calculated for the case D>0 does not agree with any of the intensity ratios derived from the experiments, whereas that for the case D<0 is in excellent agreement with the observed intensity ratio $(I_1/I_3)_{obs}=(0.47\pm0.06)$, thus the energies of these two excitations can unambiguously be identified as $\varepsilon_1$ and $\varepsilon_3$ and the anisotropy parameter D is definitely negative, so that the ground state is $|2,\frac{3}{2},\pm\frac{3}{2}>$.

The remaining transition at energy $\varepsilon_2$ therefore corresponds to the anisotropy-induced splitting of the ground state $a_2|0,1>-a_1|1,1>$ of the trimer Cu-Cu-Ni; more specifically, with D<0 the transition has to be identified as $(a_2|0,1,\pm1>-a_1|1,1,\pm1>) \rightarrow (a_2|0,1,0>-a_1|1,1,0>)$. The low-lying excitations associated with both trimers Cu-Cu-Ni and Ni-Cu-Ni are best described by the parameters

$J_{Cu-Ni} = (-0.85 \pm 0.10)$ meV , $D_{Ni} = (-0.7 \pm 0.1)$ meV .

So far we did not consider the detailed population of the different trimer types. The comparison of the observed and calculated intensities $I_1$ and $I_3$ for the trimer Ni-Cu-Ni as well as $I_2$ for the trimer Cu-Cu-Ni, however, allows to estimate these populations. The calculated intensity ratios $(I_1/I_2)_{calc}=0.38$ and $(I_2/I_3)_{calc}=1.16$ are slightly different from the observed intensity ratios $(I_1/I_2)_{obs}=(0.35\pm0.04)$ and $(I_2/I_3)_{obs}=(1.35\pm0.11)$, which can be reconciled with the assumption that the population of the trimers Cu-Cu-Ni exceeds that of the trimers Ni-Cu-Ni by 12%, resulting in trimer populations Cu-Cu-Cu / Cu-Cu-Ni / Ni-Cu-Ni of 32% / 36% / 32%, respectively. This is a surprising result, since purely statistical considerations suggest a double population of the trimers Cu-Cu-Ni (and Ni-Cu-Cu) with respect to the trimers Ni-Cu-Ni (and Cu-Cu-Cu).

Obviously the compound $Ca_3Cu_2Ni(PO_4)_4$ has a strong preference to form symmetric trimers Ni-Cu-Ni rather than asymmetric trimers Cu-Cu-Ni.

The above conclusion about the trimer populations is qualitatively confirmed in Fig. 8 by a comparison between the observed and calculated spectra; the latter correspond to a superposition of the excitations associated with the three trimer types with different trimer populations. Neither the trimer populations governed by statistics (25% / 50% / 25%) nor the "symmetry preferred" trimer populations (50% / 0% / 50%) are able to describe the observed data in a satisfactory manner. The best qualitative agreement is obtained for the trimer populations 36% / 28% / 36% which are very close to the results derived from the analysis of Fig. 7.

### 4.4. $Ca_3CuNi_2(PO_4)_4$

For the compound $Ca_3CuNi_2(PO_4)_4$ the trimers are exclusively of the type Ni-Cu-Ni, thus we can readily calculate the trimer splittings from the parameters $J_{Cu-Ni}$ and $D_{Ni}$ determined in the preceding Section. The calculation predicts the highest excited trimer state $|2,\frac{5}{2},\pm\frac{1}{2}>$ to lie at 5.10 meV, see Table 1. Since $Ca_3CuNi_2(PO_4)_4$ exhibits antiferromagnetic ordering below $T_N=20$ K [14], the measurements have to be carried out in the paramagnetic state in order to avoid a further splitting of the trimer states due to the molecular field. However, all the energy spectra taken at $T>T_N$ do not exhibit well defined inelastic lines but rather broad intensity distributions. This is due to the presence of excited-state transitions as well as to line broadening resulting from magnetic inter-trimer interactions. Therefore we were unable to carry out a rigorous analysis of the energy spectra taken in the paramagnetic state.

Fig. 9 displays an energy spectrum taken at T=4 K << $T_N$. There is a pronounced peak at $\varepsilon_1$=(6.05±0.03) meV with a shoulder on the low-energy side at $\varepsilon_2$=(4.71±0.22) meV. As schematically sketched in Fig. 3, the molecular field splits the ground-state doublet $|2,\frac{3}{2},\pm\frac{3}{2}\rangle$ in such a way that the singlet $|2,\frac{3}{2},\frac{3}{2}\rangle$ lies lowest, separated from its antisymmetric counterpart $|2,\frac{3}{2},-\frac{3}{2}\rangle$ by the energy $3g\mu_B H_{mf}$. In principle, the transitions between the molecular-field states exhibit dispersion whose wave-vector dependence can only be unravelled in single-crystal experiments. Neutron scattering data obtained for polycrystalline material correspond to a weighted integration over the whole dispersion branch providing some sort of density-of-states with maximum intensity at the boundary of the Brillouin zone where the molecular-field solution is exactly realized. Thus we identify the peaks at $\varepsilon_1$ and $\varepsilon_2$ with ground-state transitions to the two highest molecular-field states which are allowed by the selection rules (13). More specifically, the peaks at $\varepsilon_1$ and $\varepsilon_2$ are attributed to the transitions $|2,\frac{3}{2},\frac{3}{2}\rangle \rightarrow |2,\frac{5}{2},\frac{1}{2}\rangle$ and $|2,\frac{3}{2},\frac{3}{2}\rangle \rightarrow |2,\frac{5}{2},\frac{3}{2}\rangle$, respectively. According to Eq. (11) the peak at $\varepsilon_1$ is shifted upwards from the trimer state $|2,\frac{5}{2},\pm\frac{1}{2}\rangle$ at energy 5.10 meV (see Table 1) by $g\mu_B H_{mf}$, whereas the peak at $\varepsilon_2$ is expected to remain at the energy of the trimer state $|2,\frac{5}{2},\pm\frac{3}{2}\rangle$ at 4.70 meV (see Table 1). From the molecular-field shift of the former peak we find $g\mu_B H_{mf}$=(0.95±0.18) meV, corresponding to $H_{mf}$=(8.2±1.6) Tesla. The molecular-field parameter $\lambda$ can then be derived from Eq. (10) which yields $\lambda$=(0.63±0.12) meV with $\langle\mathbf{S}\rangle=\frac{3}{2}$ being the expectation value of the total spin operator $\mathbf{S}$ in the ground state.

In Fig. 9 the transitions to the remaining excited states $|1,\frac{3}{2},\pm M\rangle$ and $|0,\frac{1}{2},\pm\frac{1}{2}\rangle$ do not show up. The reasons for the absence of these transitions are twofold. Firstly, the ground-state transition to $|0,\frac{1}{2},\pm\frac{1}{2}\rangle$ is forbidden due to the broken selection rule $\Delta S_{13}$=-2 (see Eq. 13). Secondly, the neutron cross-section

for ground-state transitions to $|2,\frac{5}{2},\pm M>$ exceeds that for $|1,\frac{3}{2},\pm M>$ by almost an order of magnitude, thus the latter does not significantly contribute to the scattering.

## 5. Discussion and conclusions

By analyzing the inelastic neutron scattering data observed for $Ca_3Cu_{3-x}Ni_x(PO_4)_4$ (x=0,1,2) we were able to determine the relevant interaction parameters associated with the Cu-Cu-Cu, Cu-Cu-Ni, and Ni-Cu-Ni trimers in a consistent manner. The results of our analyses are summarized in Table 1. The Cu-Cu-Cu and trimers are defined by pure $|S_{13},S,\pm M>$ states, whereas for the Cu-Cu-Ni and Ni-Cu-Ni trimers a mixing of the states with equal spin quantum number M occurs. However, with $|J_{Cu-Ni}|<<|J_{Cu-Cu}|$ and $|D_{Ni}|<<|J_{Cu-Cu}|$ the M-mixing is modest, i.e., for all mixed states $a_1|S_{13},S,\pm M>+a_2|S'_{13},S,\pm M>$ we have either $a_1<<a_2$ or $a_1>>a_2$, thus the states of the Cu-Cu-Ni and Ni-Cu-Ni trimers listed in Table 1 are described by the dominating substate $|S_{13},S,\pm M>$ alone.

The most important physical information resulting from our study concerns the definition of the ground state of the three trimer types which should be compatible with the spin configuration as well as with the magnetic saturation moments in the magnetically ordered state. High-field magnetization measurements performed for the Cu-Cu-Cu trimer system $Ca_3Cu_3Ni(PO_4)_4$ just above the ordering temperature show a saturation at $\mu_{Cu}=1.15$ $\mu_B$ for fields exceeding 10 T [12] which is compatible with the S=1/2 ground state $|1,\frac{1}{2}>$ determined in the present work. The situation is more complex for the compound $Ca_3CuNi_2(PO_4)_4$. Although our analysis gave evidence for an antiferromagnetic intratrimer exchange coupling $J_{Cu-Ni}$, the presence of two different sublattices a and b with antiferromagnetic and ferromagnetic intratrimer spin alignment, respectively, were reported from neutron diffraction [14]. Obviously the trimer model alone cannot explain the existence of the ferromagnetically aligned trimer sublattice, but intertrimer interactions (which are of the order of $\lambda$=0.65 meV and therefore comparable to $J_{Cu-Ni}$=-0.85 meV) have to be considered. The latter may be responsible for the considerable quenching of the $Cu^{2+}$ moment as well. The ground state $|2,\frac{3}{2},\pm\frac{3}{2}>$ determined in the present work confirms the negative sign of the anisotropy parameter $D_{Ni}$, since for $D_{Ni}$>0 the state $|2,\frac{3}{2},\pm\frac{1}{2}>$ would be the ground state, so that also the $Ni^{2+}$ moments would be strongly quenched.

By analyzing the energy spectra associated with the Ni-Cu-Ni trimers in the magnetically ordered state of the compound $Ca_3CuNi_2(PO_4)_4$ we were able to determine the molecular-field parameter $\lambda$ which is related to the Néel temperature by

$$T_N = \tfrac{2}{3}S(S+1)\lambda \,. \tag{14}$$

Inserting $\lambda$=(0.63±0.12) meV and S=$\tfrac{3}{2}$ yields $T_N$=(18.3±3.5) K which is in good agreement with the observed ordering temperature $T_N$=20 K [14].

The size of the exchange parameter $J_{Cu-Cu}$ of the compounds $Ca_3Cu_{3-x}Ni_x(PO_4)_4$ was found to increase by about 4% when going from x=0 to x=1. Superexchange between the $Cu^{2+}$ ions mediated through the bridging oxygen is clearly the dominating exchange mechanism in the title compounds. From neutron diffraction the following bond distances R and bond angles $\Theta$ were reported [14]:

x=0: $R_{Cu1-O}$=1.917(2) Å , $R_{Cu2-O}$=2.086(4) Å , $R_{Cu1-Cu2}$=3.5334(19) Å ,
$\Theta_{Cu1-O-Cu2}$=123.88(15)° ;

x=1: $R_{Cu1-O}$=1.953(2) Å , $R_{Cu2-O}$=2.099(4) Å , $R_{Cu1-Cu1}$=3.5583(19) Å ,
$\Theta_{Cu1-O-Cu2}$=124.51(17)° .

Both R and $\Theta$ are increasing for increasing Ni content. Clearly these bond characteristics are in competition, since usually a decrease of R and an increase of Θ results in an increase of |J|. Obviously the bond angle Θ has a stronger influence on J in the title compounds.

The compound $Ca_3Cu_2Ni(PO_4)_4$ is of particular interest due to the simultaneous presence of Cu-Cu-Cu, Cu-Cu-Ni, and Ni-Cu-Ni trimers whose populations, however, cannot be assessed by diffraction experiments alone. The analyses of our neutron spectroscopic data show that the trimer populations deviate significantly from those predicted by purely statistical considerations. More specifically, "nature" has a strong preference to favor symmetric Cu-Cu-Cu and Ni-Cu-Ni trimers rather than asymmetric Cu-Cu-Ni trimers.

In conclusion, we have shown, using neutron spectroscopic measurements on polycrystalline materials, that the dynamic magnetic properties of the compounds $Ca_3Cu_{3-x}Ni_x(PO_4)_4$ are essentially determined by the low-lying electronic states of the built-in Cu-Cu-Cu, Cu-Cu-Ni, and Ni-Cu-Ni trimers. We have been able to parametrize the relevant trimer interactions in terms of nearest-neighbor exchange and axial anisotropy, whereas next-nearest-neighbor exchange was found to be negligibly small. Our work forms a useful basis for the further study of quantum magnetism in (mixed) spin-trimer compounds.

**Acknowledgments**

This work was performed at the Swiss Spallation Neutron Source SINQ, Paul Scherrer Institut (PSI), Villigen, Switzerland. Financial support by the NCCR MaNEP project is gratefully acknowledged.

**Table Captions**

**Table 1:**

Spin states and excitation energies of Cu-Cu-Cu, Cu-Cu-Ni, and Ni-Cu-Ni trimers in $Ca_3Cu_{2-x}Ni_x(PO_4)_4$ (x=0,1,2) at T≤4 K.

a) Cu-Cu-Cu trimers in $Ca_3Cu_{2-x}Ni_x(PO_4)_4$.

b) Ni-Cu-Ni trimers in $Ca_3Cu_{2-x}Ni_x(PO_4)_4$. $J_{Cu-Ni}$=-0.85(10) meV, $D_{Ni}$=-0.7(1) meV.

c) Cu-Cu-Ni trimers in $Ca_3Cu_2Ni(PO_4)_4$. $J_{Cu-Cu}$=-4.92(5) meV, $J_{Cu-Ni}$=-0.85(10) meV, $D_{Ni}$=-0.7(1) meV.

**Figure Captions**

**Figure 1:**

Possible types of spin trimers relevant for the compounds $Ca_3Cu_{3-x}Ni_x(PO_4)_4$.

**Figure 2:**

Spin state correlation diagram for a mixed trimer of type III.A with $A=Cu^{2+}$ and $B=Ni^{2+}$.

**Figure 3:**

Schematic sketch of the effect of a magnetic field on trimer states $|S,M\rangle$ (the spin quantum number $S_{13}$ is omitted).

**Figure 4:**

Q-dependence of the neutron cross-section for ground-state transitions in an antiferromagnetically coupled spin trimer of type I.A with $A=Cu^{2+}$ and R=3.56 Å. The circles denote the intensities observed for the compound $Ca_3Cu_3(PO_4)_4$. The intensities are normalized for the $|1,\frac{1}{2}\rangle \rightarrow |1,\frac{3}{2}\rangle$ transition at $\langle Q \rangle = 2$ Å$^{-1}$.

**Figure 5:**

Energy spectra of neutrons scattered from $Ca_3Cu_{3-x}Ni_x(PO_4)_4$ (x=0,1,2) at T=1.5 K, $\langle Q \rangle$=2.7 Å$^{-1}$, and $E_i$=20 meV. The data for x=0 and x=1 are shifted upwards by 13 and 5 units, respectively.

**Figure 6:**

Energy spectra of neutrons scattered from $Ca_3Cu_3(PO_4)_4$ at T=1.5 K and $E_i$=20 meV for different moduli of the scattering vector **Q**. The data for <Q> = 2.25 and 2.85 $Å^{-1}$ are shifted upwards by 12 and 6 units, respectively.

**Figure 7:**

Low-energy spectrum of neutrons scattered from $Ca_3Cu_2Ni(PO_4)_4$ at T=1.5 K, <Q>=1.1 $Å^{-1}$, and $E_i$=2.7 meV. The lines correspond to the results of a least-squares fit as described in the text.

**Figure 8:**

Energy spectrum of neutrons scattered from $Ca_3Cu_2Ni(PO_4)_4$ at T=1.5 K, <Q>=2.7 $Å^{-1}$, and $E_i$=20 meV. The lines correspond to the calculated spectra which are the result of a superposition of the excitations associated with the trimers Cu-Cu-Cu/Cu-Cu-Ni/Ni-Cu-Ni for different trimer populations as indicated in the Figure. The calculated data are shifted upwards by 4 units.

**Figure 9:**

Energy spectrum of neutrons scattered from $Ca_3CuNi_2(PO_4)_4$ at T=4 K, <Q>=1.9 $Å^{-1}$ and $E_i$=15 meV. The lines correspond to the results of a Gaussian least-squares fit.

**Table 1:**
Spin states and excitation energies of Cu-Cu-Cu, Cu-Cu-Ni, and Ni-Cu-Ni trimers in $Ca_3Cu_{2-x}Ni_x(PO_4)_4$ (x=0,1,2) at T≤4 K.

a) Cu-Cu-Cu trimers in $Ca_3Cu_{2-x}Ni_x(PO_4)_4$.

| $|S_{13},S\rangle$ | x=0: $J_{Cu-Cu}$=-4.74(2) meV, $J'_{Cu-Cu}$=-0.02(3) meV | | x=1: $J_{Cu-Cu}$=-4.92(5) meV, $J'_{Cu-Cu}$=0 | |
|---|---|---|---|---|
| | $E_{calc}$ [meV] | $E_{obs}$ [meV] | $E_{calc}$ [meV] | $E_{obs}$ [meV] |
| $|1,1/2\rangle$ | 0 | 0 | 0 | 0 |
| $|0,1/2\rangle$ | 9.44 | 9.44 ± 0.03 | 9.84 | - |
| $|1,3/2\rangle$ | 14.22 | 14.22 ± 0.02 | 14.76 | 14.76 ± 0.14 |

b) Ni-Cu-Ni trimers in $Ca_3Cu_{2-x}Ni_x(PO_4)_4$. $J_{Cu-Ni}$=-0.85(10) meV, $D_{Ni}$=-0.7(1) meV.

| $|S_{13},S,\pm M\rangle$ | $E_{calc}$ [meV] | $E_{obs}$ [meV] (x=1) | $E_{obs}$ [meV] (x=2) |
|---|---|---|---|
| $|2,3/2,\pm 3/2\rangle$ | 0 | 0 | 0 |
| $|2,3/2,\pm 1/2\rangle$ | 0.59 | 0.52 ± 0.06 | - |
| $|1,1/2,\pm 1/2\rangle$ | 1.15 | 1.05 ± 0.06 | - |
| $|0,1/2,\pm 1/2\rangle$ | 2.84 | - | - |
| $|1,3/2,\pm 1/2\rangle$ | 3.56 | - | - |
| $|1,3/2,\pm 3/2\rangle$ | 3.98 | - | - |
| $|2,5/2,\pm 5/2\rangle$ | 4.13 | - | - |
| $|2,5/2,\pm 3/2\rangle$ | 4.70 | - | 4.71 ± 0.22 |
| $|2,5/2,\pm 1/2\rangle$ | 5.10 | - | - |

c) Cu-Cu-Ni trimers in $Ca_3Cu_2Ni(PO_4)_4$. $J_{Cu-Cu}$=-4.92(5) meV, $J_{Cu-Ni}$=-0.85(10) meV, $D_{Ni}$=-0.7(1) meV.

| $|S_{12},S,\pm M\rangle$ | $E_{calc}$ [meV] | $E_{obs}$ [meV] |
|---|---|---|
| $|0,1,\pm 1\rangle$ | 0 | 0 |
| $|0,1,0\rangle$ | 0.68 | 0.80 ± 0.04 |
| $|1,0,0\rangle$ | 7.88 | - |
| $|1,1,0\rangle$ | 9.31 | - |
| $|1,1,\pm 1\rangle$ | 9.57 | - |
| $|1,2,\pm 2\rangle$ | 10.84 | - |
| $|1,2,\pm 1\rangle$ | 11.27 | - |

| $|1,2,0\rangle$ | 11.95 | - |
---

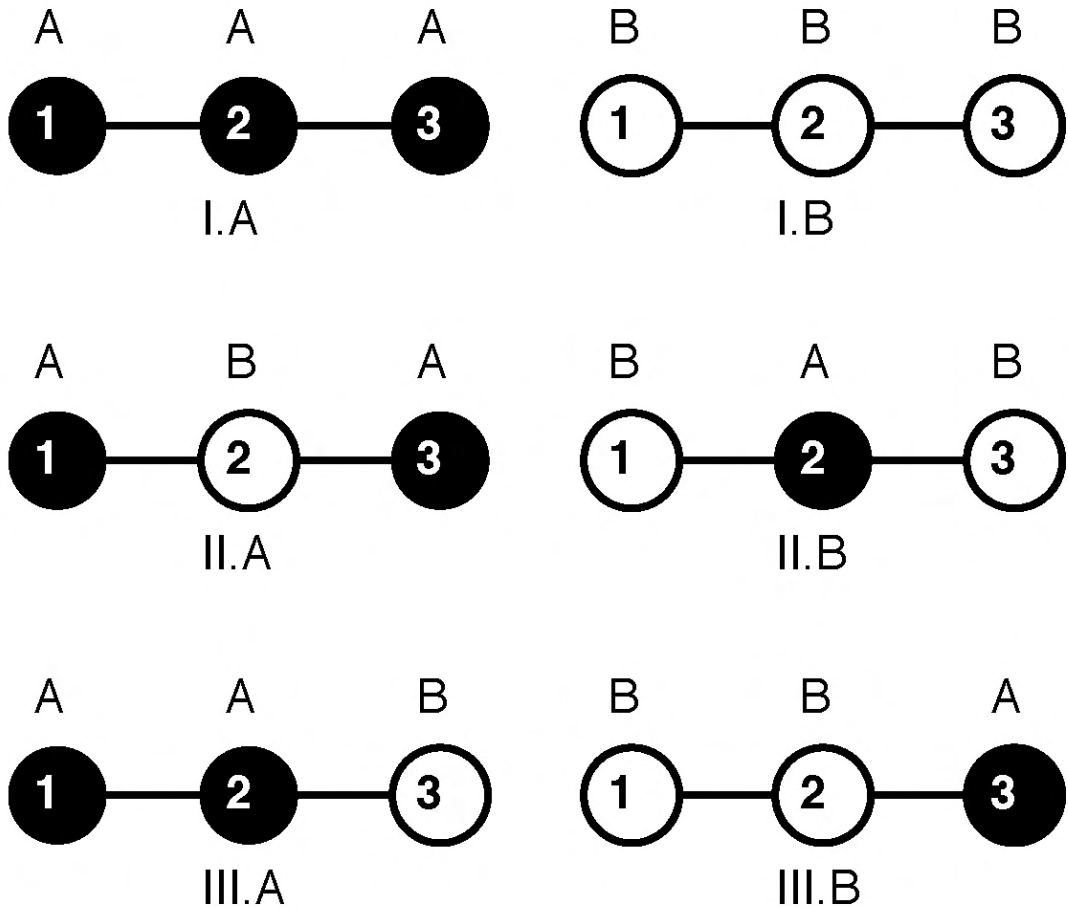

**Figure 1:**

Possible types of spin trimers relevant for the compounds $Ca_3Cu_{3-x}Ni_x(PO_4)_4$.

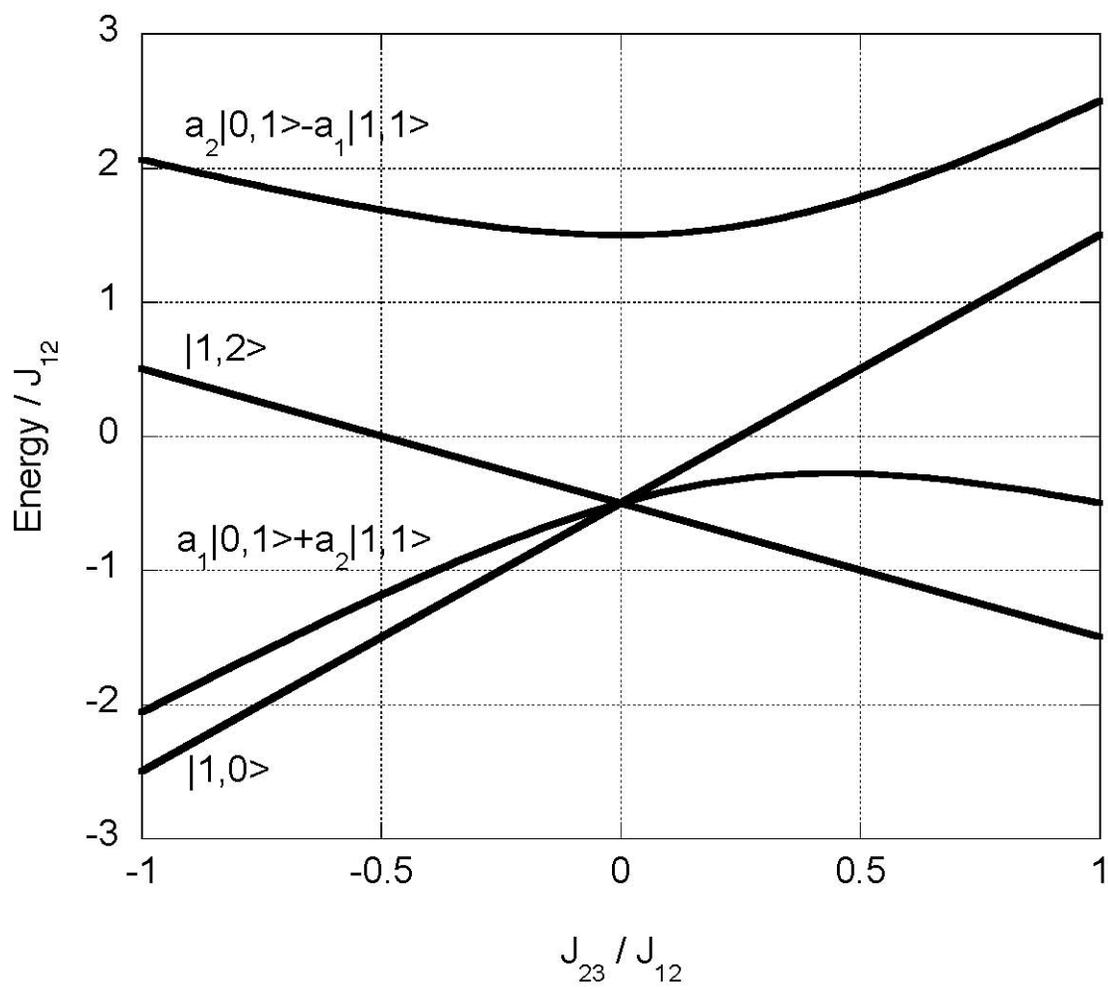

**Figure 2:**

Spin state correlation diagram for a mixed trimer of type III.A with A=$Cu^{2+}$ and B=$Ni^{2+}$.

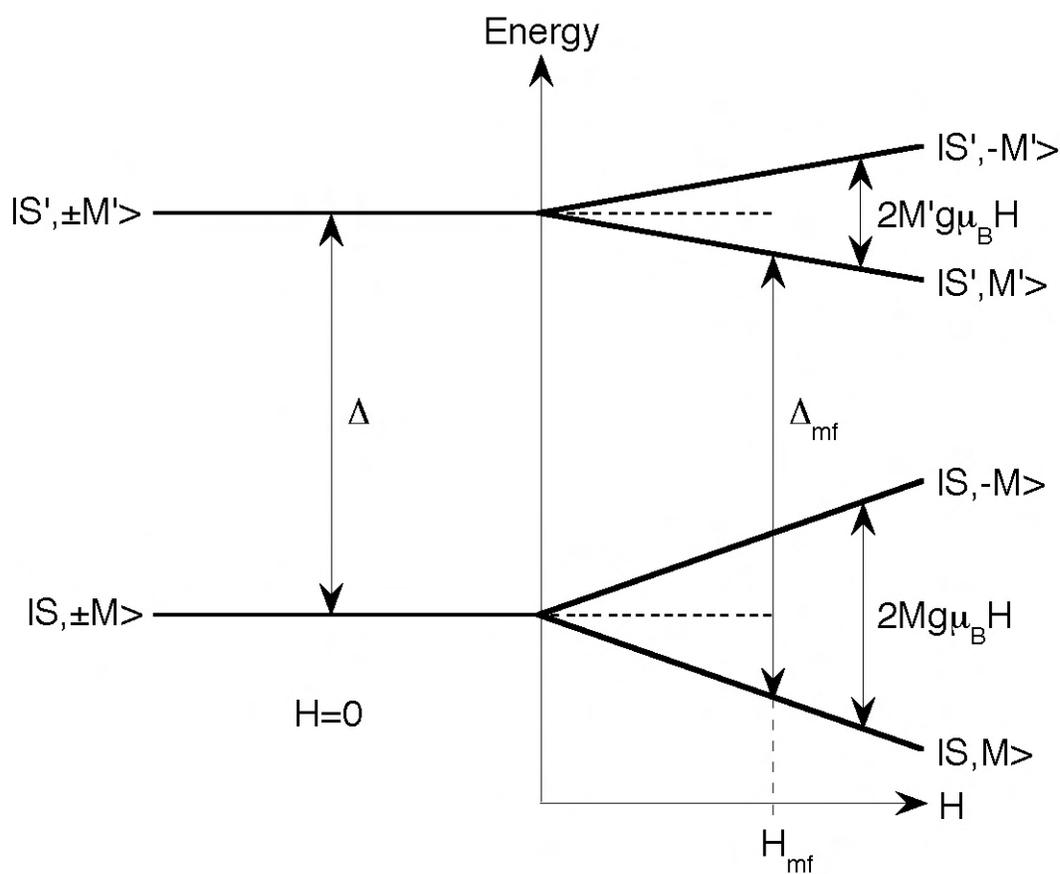

**Figure 3:**

Schematic sketch of the effect of a magnetic field on trimer states |S,M> (the spin quantum number $S_{13}$ is omitted).

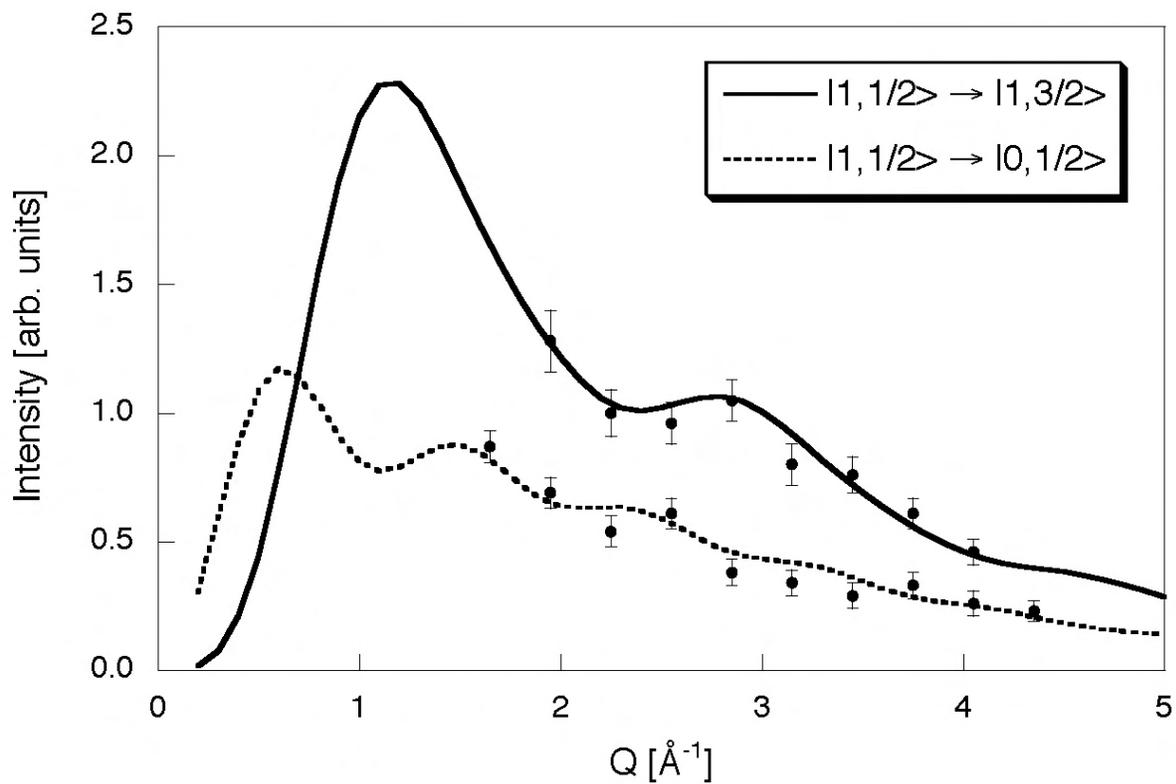

**Figure 4:**

Q-dependence of the neutron cross-section for ground-state transitions in an antiferromagnetically coupled spin trimer of type I.A with A=$Cu^{2+}$ and R=3.56 Å. The circles denote the intensities observed for the compound $Ca_3Cu_3(PO_4)_4$. The intensities are normalized for the $|1,\frac{1}{2}>\rightarrow|1,\frac{3}{2}>$ transition at $<Q> = 2$ Å$^{-1}$.

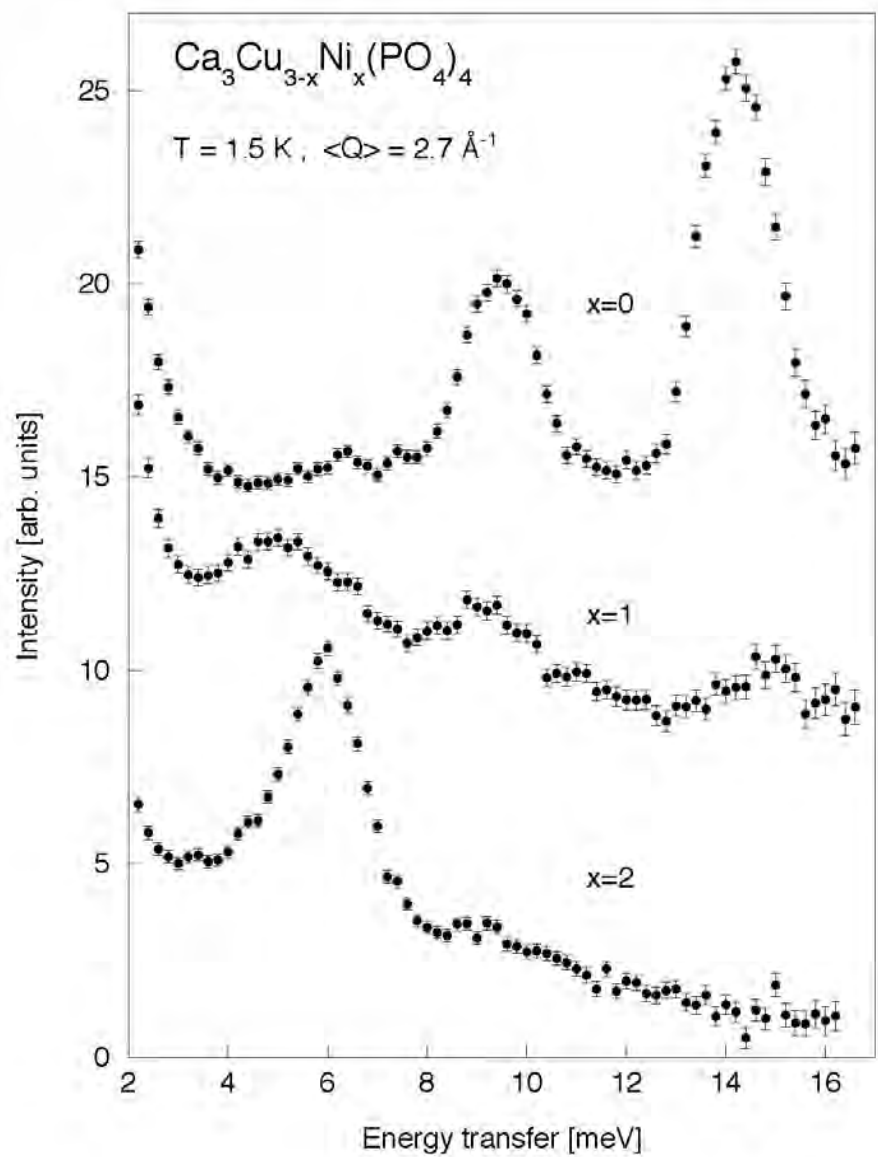

**Figure 5:**

Energy spectra of neutrons scattered from $Ca_3Cu_{3-x}Ni_x(PO_4)_4$ (x=0,1,2) at T=1.5 K, <Q>=2.7 Å$^{-1}$, and $E_i$=20 meV. The data for x=0 and x=1 are shifted upwards by 13 and 5 units, respectively.

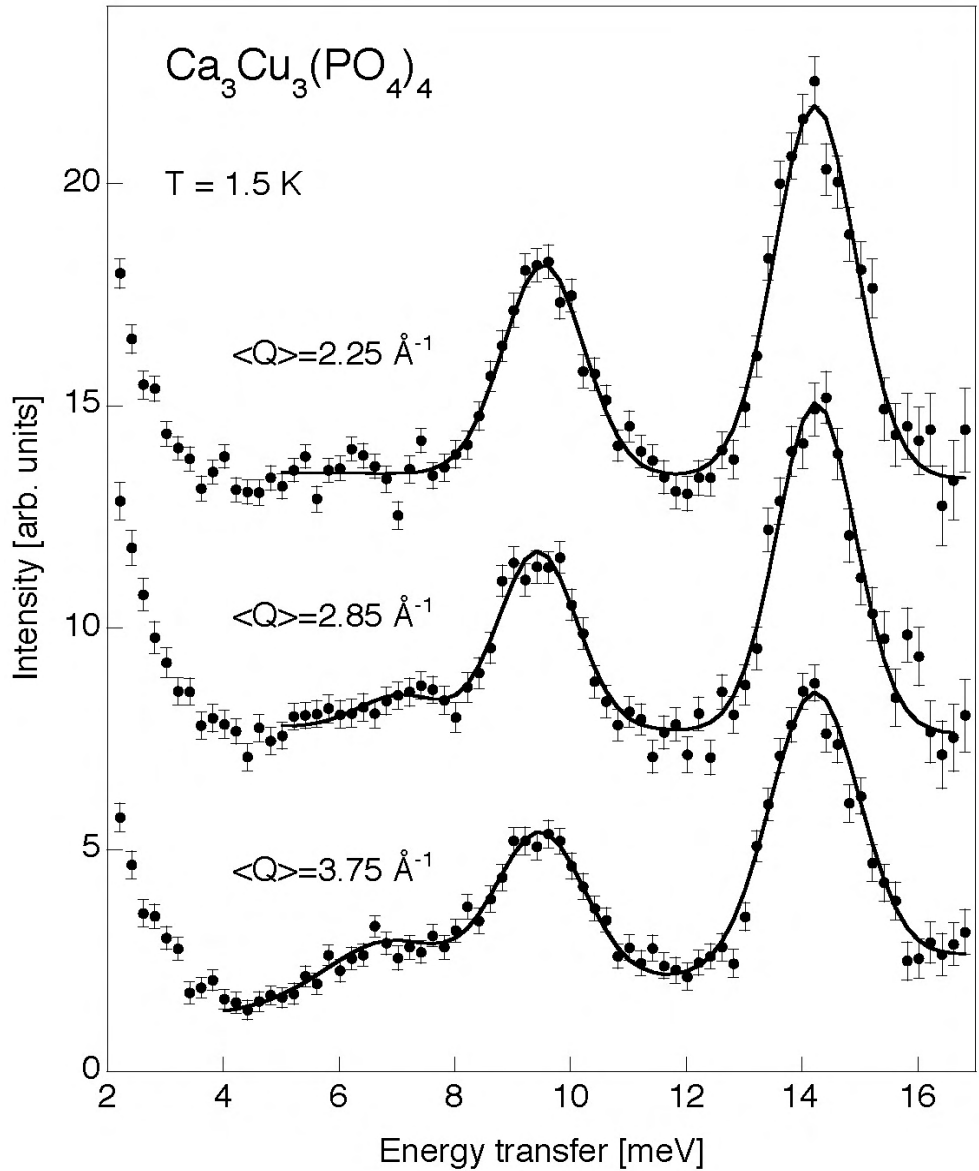

**Figure 6:**

Energy spectra of neutrons scattered from $Ca_3Cu_3(PO_4)_4$ at T=1.5 K and $E_i$=20 meV for different moduli of the scattering vector **Q**. The data for <Q> = 2.25 and 2.85 Å$^{-1}$ are shifted upwards by 12 and 6 units, respectively.

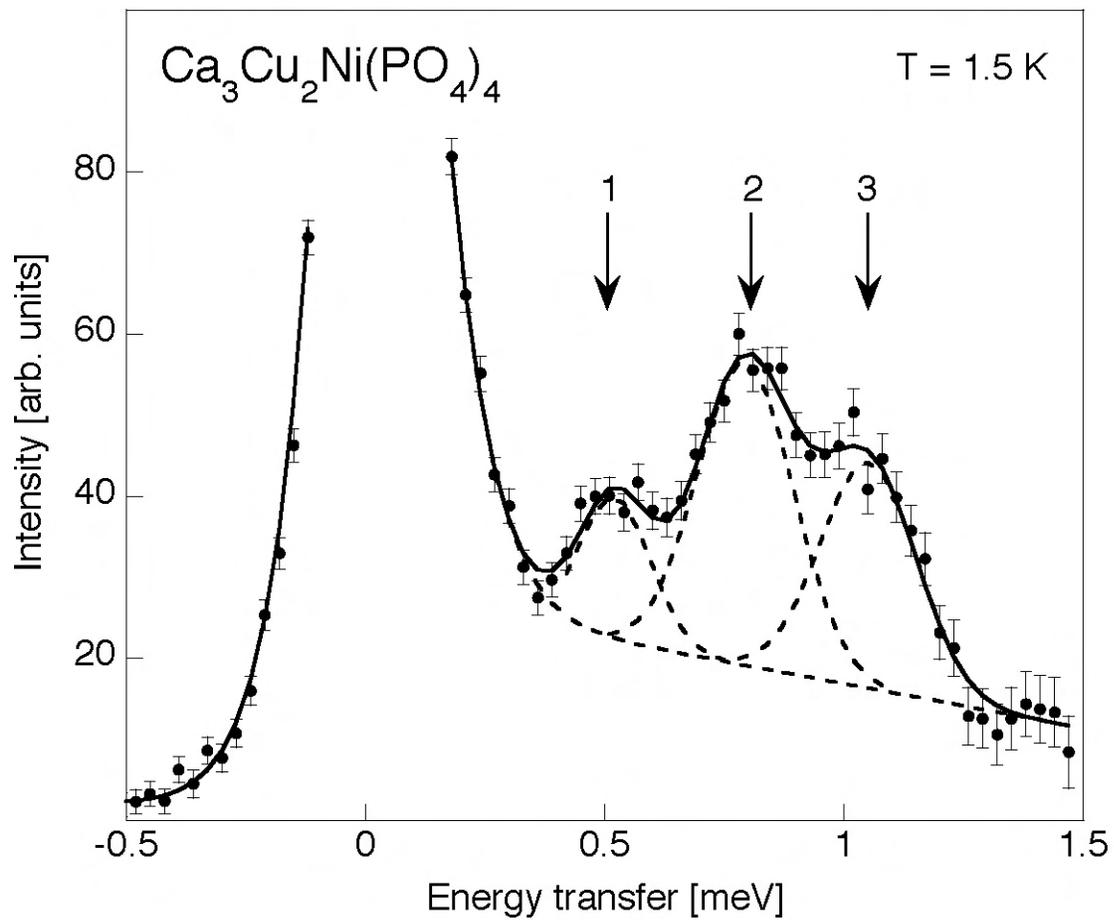

**Figure 7:**

Low-energy spectrum of neutrons scattered from $Ca_3Cu_2Ni(PO_4)_4$ at T=1.5 K, $<Q>$=1.1 Å$^{-1}$, and $E_i$=2.7 meV. The lines correspond to the results of a least-squares fit as described in the text.

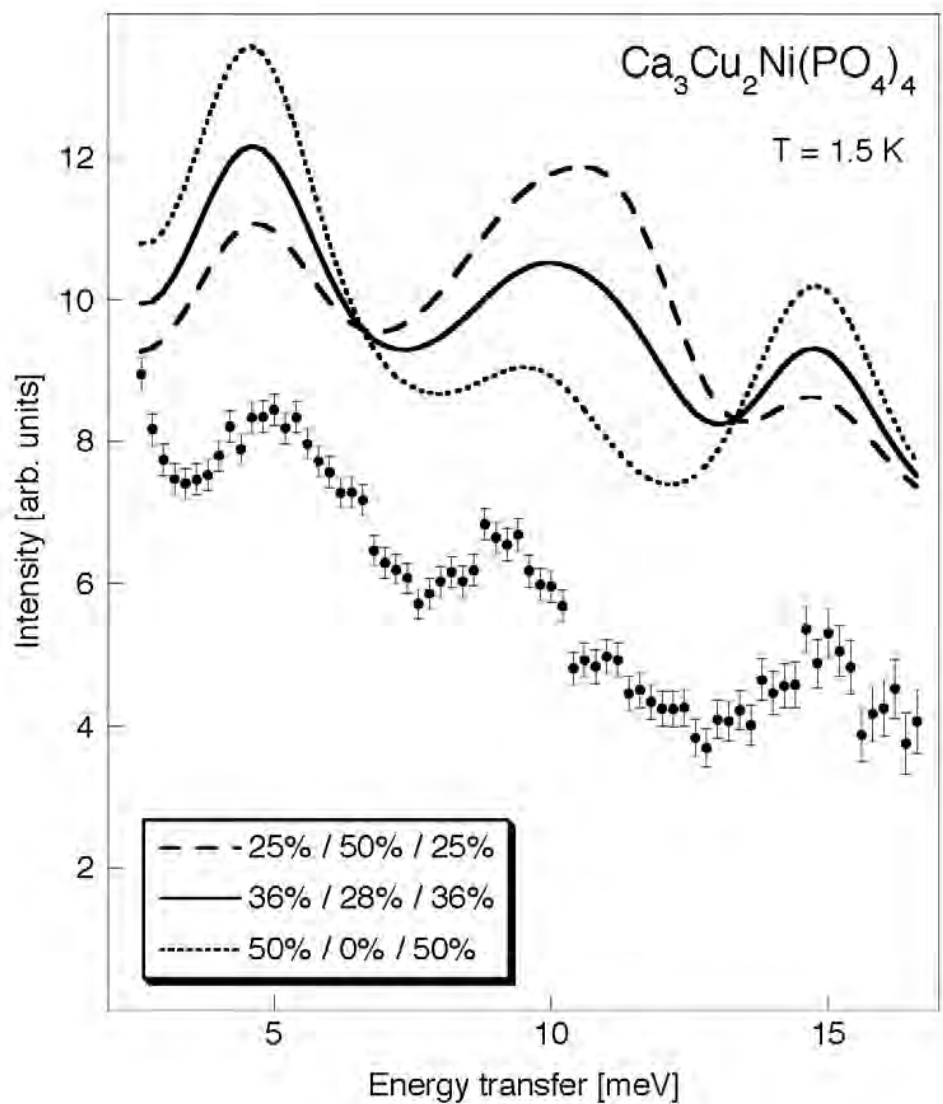

**Figure 8:**

Energy spectrum of neutrons scattered from $Ca_3Cu_2Ni(PO_4)_4$ at T=1.5 K, $<Q>$=2.7 Å$^{-1}$, and $E_i$=20 meV. The lines correspond to the calculated spectra which are the result of a superposition of the excitations associated with the trimers Cu-Cu-Cu/Cu-Cu-Ni/Ni-Cu-Ni for different trimer populations as indicated in the Figure. The calculated data are shifted upwards by 4 units.

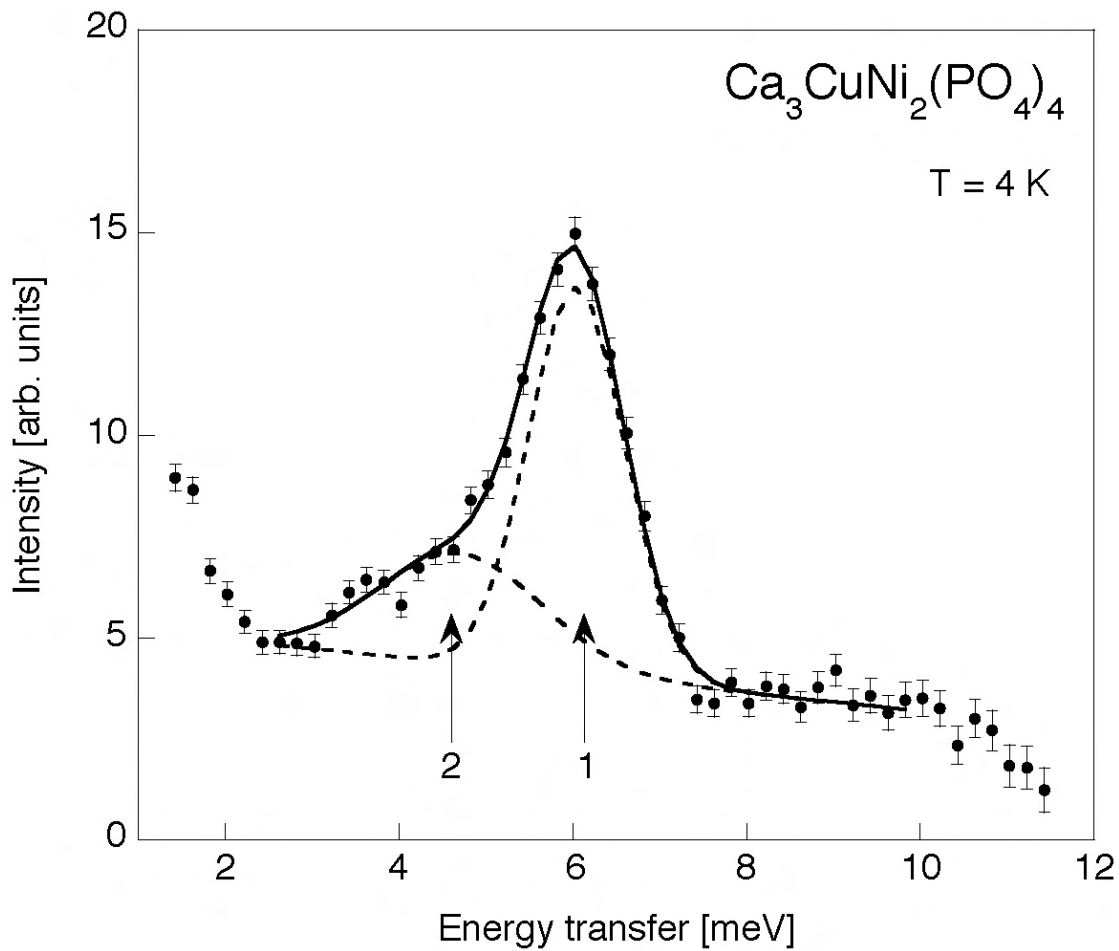

**Figure 9:**

Energy spectrum of neutrons scattered from $Ca_3CuNi_2(PO_4)_4$ at T=4 K, $<Q>$=1.9 Å$^{-1}$ and $E_i$=15 meV. The lines correspond to the results of a Gaussian least-squares fit.